\documentclass[twocolumn]{aastex62}
\usepackage{amsmath}
\graphicspath{{./}{figures/}}

\shorttitle{Discovery of a Meteor of Interstellar Origin}

\begin{document}

\title{Discovery of a Meteor of Interstellar Origin}

\email{amir.siraj@cfa.harvard.edu, aloeb@cfa.harvard.edu}

\author{Amir Siraj}
\affil{Department of Astronomy, Harvard University, 60 Garden Street, Cambridge, MA 02138, USA}

\author{Abraham Loeb}
\affiliation{Department of Astronomy, Harvard University, 60 Garden Street, Cambridge, MA 02138, USA}




\begin{abstract}
The earliest confirmed interstellar object, `Oumuamua, was discovered in the Solar System by Pan-STARRS in 2017, allowing for a calibration of the abundance of interstellar objects of its size $\sim 100\;$ m. This was followed by the discovery of Borisov, which allowed for a similar calibration of its size $\sim 0.4 - 1 \mathrm{\; km}$. One would expect a much higher abundance of significantly smaller interstellar objects, with some of them colliding with Earth frequently enough to be noticeable. Based on the CNEOS catalog of bolide events, we identify the $\sim 0.45$m meteor detected at 2014-01-08 17:05:34 UTC as originating from an unbound hyperbolic orbit with 99.999\% confidence. The U.S. Department of Defense has since verified that ``the velocity estimate reported to NASA is sufficiently accurate to indicate an interstellar trajectory.'' We infer that the meteor had an asymptotic speed of $v_{\infty} \sim 42.1 \pm 5.5\; \mathrm{km \; s^{-1}}$ outside of the solar system. Its origin is approximately towards R.A. 
$49.4 \pm 4.1^{\circ}$ and declination $11.2 \pm 1.8^{\circ}$, implying that its initial velocity vector was $58\pm6\; \mathrm{km\;s^{-1}}$ away from the velocity of the Local Standard of Rest (LSR). Its high LSR speed implies a possible origin from the deep interior of a planetary system or a star in the thick disk of the Milky Way galaxy. The local number density of its population is $10^{6{^{+0.75}_{-1.5}}} \; \mathrm{AU^{-3}}$ or $9 \times 10^{21{^{+0.75}_{-1.5}}} \; \mathrm{pc^{-3}}$ (necessitating 0.2 -- 20 Earth masses of material to be ejected per local star). We show that the detections of CNEOS 2014-01-08, `Oumuamua, and Borisov collectively imply that the differential size distribution in good agreement with a collisional distribution, with a power-law slope is $q \sim 3.6 \pm 0.5$, where the quoted uncertainty corresponds to $2 \sigma$. We then consider the possibility of analyzing interstellar meteor compositions based on spectroscopy of their gaseous debris as they burn up in the Earth's atmosphere. We propose a strategy for determining the orbits and chemical compositions of interstellar meteors, using a network of $\sim 600$ all-sky camera systems to track and conduct remote spectroscopy on meteors larger than $\sim 5$cm once every few years. It should also be possible to retrieve meteorites from the impact sites, providing the first samples of materials from other planetary systems.

\end{abstract}



\section{Introduction}
`Oumuamua was the first interstellar object detected in the Solar System by Pan-STARRS \citep{Meech2017, Micheli2018}. Several follow-up studies of `Oumuamua were conducted to better understand its origin and composition \citep{Bannister2017, Gaidos2017, Jewitt2017, Mamajek2017, Ye2017, Bolin2017, Fitzsimmons2018, Trilling2018, Bialy2018, Hoang2018, Siraj2019a, Siraj2019b, Seligman2019}. Its size was estimated to be 20m -- 200m, based on Spitzer Space Telescope constraints on its infrared emission given its temperature \citep{Trilling2018}. The discovery of `Oumuamua was followed by that of the second interstellar object, Borisov, in 2019 \citep{2020NatAs...4...53G}. The size of Borisov's nucleus was estimated to be $0.4 - 1 \mathrm{\; km}$ \citep{2020ApJ...888L..23J, 2021MNRAS.507L..16S}. 

\cite{Forbes2019} predicted that spectroscopy of `Oumuamua-like objects grazing the Sun could reveal their chemical compositions. Since there should be a higher abundance of interstellar objects smaller than `Oumuamua, we could observe small interstellar objects impacting the Earth's atmosphere. Spectroscopy of the gaseous debris from such objects as they burn up in the Earth's atmosphere could reveal their composition. There is significant evidence for previous detections of dust-sized interstellar meteors, but to date no definitive evidence of any meter-scale interstellar meteors \citep{Opik1950, Baggaley1993, Hajdukova1994, Taylor1996, Baggaley2000, Mathews1998, Meisel2002a, Meisel2002b, Weryk2004, Afanasiev2007, Musci2012, Engelhardt2017, Hajdukova2018, 2020P&SS..19004980F}.

The CNEOS catalog includes the geocentric velocity components and geographic coordinates for bolides detected by U.S. government sensors.\footnote{https://cneos.jpl.nasa.gov/fireballs/} In this \textit{Letter}, we identify a meteor from the CNEOS catalog that is likely of interstellar origin. Furthermore,  we explore the size distribution of interstellar meteors and motivate the investigation of interstellar meteors as a new branch of astronomical research. We present a strategy for conducting spectroscopy and obtaining physical samples of interstellar meteors.

\section{Methods}
\label{sec:methods}

We analyzed the bolide events in the CNEOS catalog, and found that the meteor detected at 2014-01-08 17:05:34 UTC had an unusually high heliocentric velocity at impact.\footnote{The fastest meteor in the CNEOS catalog obtains its high speed from a head-on orbit relative to the Earth and its extrapolated orbit is found to be bound to the Sun. The meteor we focus on is the second fastest. The orbit of the third fastest meteor in the catalog is possibly bound within uncertainties.}  Accounting for the motion of the Earth relative to the Sun and the motion of the meteor relative to the Earth, we found that the meteor had a heliocentric velocity of $\sim 60\; \mathrm{km\;s^{-1}}$ at impact, which implies that the object was unbound. To uncover the kinematic history of this meteor, we integrated its motion from impact backward in time. 

The Python code created for this work used the open-source N-body integator software \texttt{REBOUND}\footnote{https://rebound.readthedocs.io/en/latest/} to trace the motion of the meteor under the gravitational influence of the Solar System \citep{Rein2012}.

We initialize the simulation with the Sun, the eight planets, and the meteor, with geocentric velocity vector $(vx_{obs}, vy_{obs}, vz_{obs}) = (-3.4, -43.5, -10.3) \; \mathrm{km\;s^{-1}}$, located at $1.3^{\circ}$ S $147.6^{\circ}$ E, at an altitude of 18.7 km, at the time of impact, $t_i =$ 2014-01-08 17:05:34 UTC, as reported in the CNEOS catalog. We then use the IAS15 adaptive time-step integrator to trace the meteor's motion back in time \citep{Rein2014}.

\section{Results}
\label{sec:results}

\subsection{Trajectory}
\label{sec:trajectory}
There are no substantial gravitational interactions between the meteor and any planet other than Earth for any trajectory within the reported errors. Based on the impact speed reported by CNEOS, $v_{obs} = 44.8\; \mathrm{km\;s^{-1}}$, we find that the meteor was unbound with an asymptotic speed of $v_{\infty} \sim 42.1\; \mathrm{km\;s^{-1}}$ outside of the solar system. In order for the object to be bound, the observed speed of $v_{obs} = 44.8 \; \mathrm{km\;s^{-1}}$ would have to be off by more than 45\%, or $20 \mathrm{\; km \;s^{-1}}$, or assuming a correct speed, a radiant off by more than $60^{\circ}$ \citep{Zuluaga2019}.

The one sigma uncertainties on each of the velocity components are better than $\pm10\%$ (M. Heavner (LANL), private communication, May 2019). Given the geocentric velocity vector $(vx_{obs}, vy_{obs}, vz_{obs}) = (-3.4 \pm 0.34, -43.5 \pm 4.35, -10.3 \pm 1.03) \; \mathrm{km\;s^{-1}}$ and assuming Gaussian statistics, we construct a trivariate normal distribution for the geocentric velocity of the meteor, centered at $(vx_{obs}, vy_{obs}, vz_{obs}) = (-3.4, -43.5, -10.3) \; \mathrm{km\;s^{-1}}$ and with root-mean-square uncertainties $(\sigma_x, \sigma_y, \sigma_z) = (0.34, 4.35, 1.03) \; \mathrm{km\;s^{-1}}$. We also construct a corresponding distribution of heliocentric impact speeds, and integrate over the probability space where the heliocentric impact speeds would imply a bound origin, $v \leq 43.6\; \mathrm{km\;s^{-1}}$. The resulting probability that an object with the measured velocities of the 2014-01-08 meteor was actually bound is $5.5 \times 10^{-8}$. As a result, we would expect $10^{-5}$ of the $\sim 270$ meteors with reported speeds in the CNEOS catalog to be bound objects with incorrectly measured velocities that imply interstellar origins as definitively as the 2014-01-08 meteor. Assuming Poisson statistics for a singular detection, we conclude with 99.999\% confidence that the 2014-01-08 meteor was interstellar.

Recently, the U.S. Department of Defense, which houses the classified data pertaining to the uncertainties involved in the CNEOS 2014-01-08 detection, released a public statement dated March 1, 2022, and addressed to the NASA Science Mission Directorate, referencing this discovery preprint and mentioning our analysis that the meteor originated from an unbound hyperbolic orbit with 99.999\% confidence \citep{Shaw2022}. The letter then states: ``Dr. Joel Mozer, the Chief Scientist of Space Operations Command, reviewed analysis of additional data available to the Department of Defense related to this finding. Dr. Mozer confirmed that the velocity estimate reported to NASA is sufficiently accurate to indicate an interstellar trajectory'' \citep{Shaw2022}. The statement thereby confirms interstellar origin of the 2014-01-08 meteor.

\begin{figure}
  \centering
  \includegraphics[width=.9\linewidth]{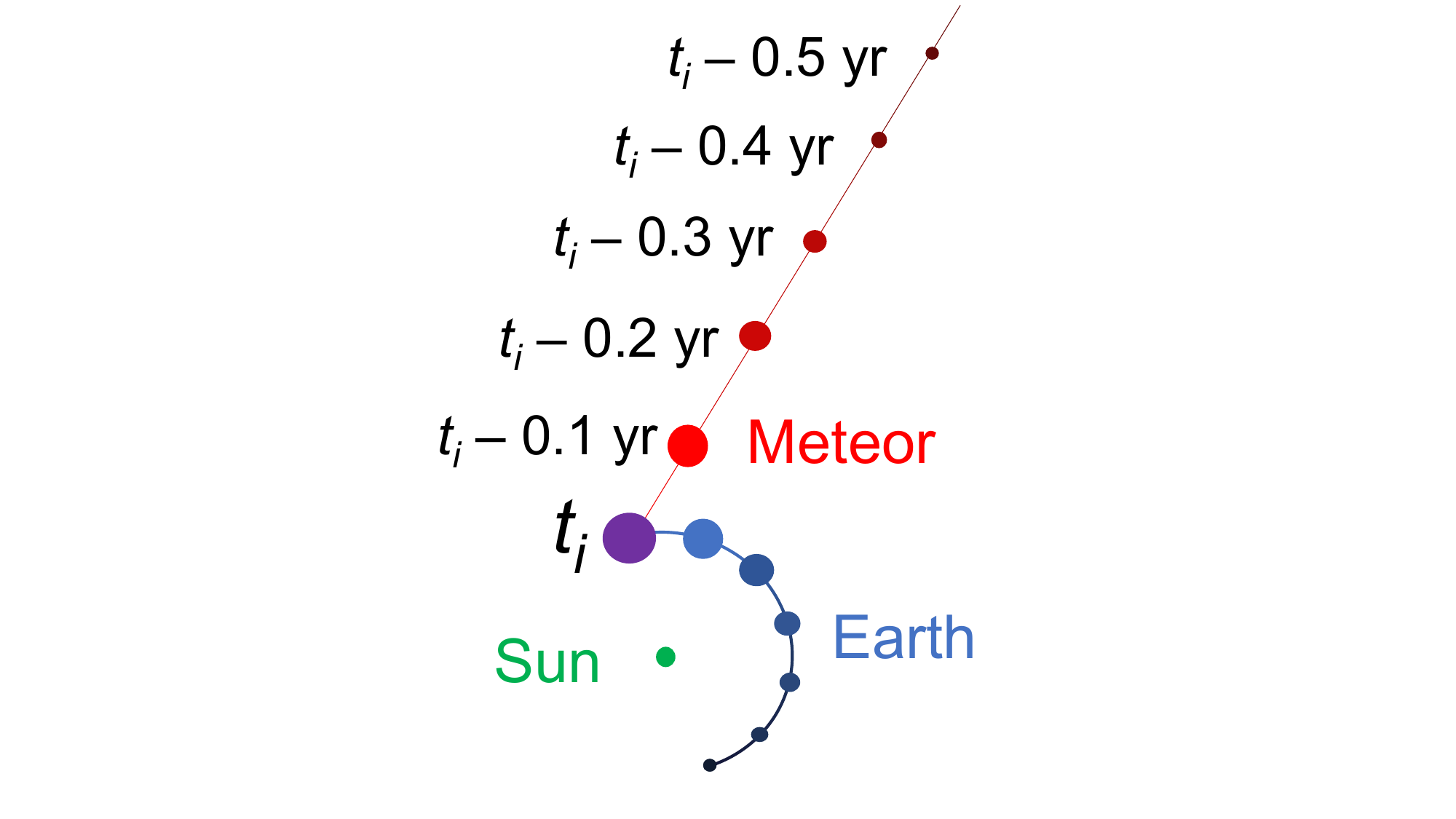}
    \caption{Trajectory of the January 8, 2014 meteor (red), shown intersecting with that of Earth (blue) at the time of impact, $t_i =$ 2014-01-08 17:05:34.}
    \label{fig:trajectory}
\end{figure}

We find that the heliocentric orbital elements of the meteor at time of impact are as follows: semi-major axis, $a = -0.47 \pm 0.15 \;$ AU, eccentricity, $e = 2.4 \pm 0.3$, inclination $i = 10\pm2^{\circ}$, longitude of the ascending node, $\Omega = 108\pm1^{\circ}$, argument of periapsis, $\omega = 58\pm2^{\circ}$, and true anomaly, $f = -58\pm2^{\circ}$. The trajectory is shown in Fig.~\ref{fig:trajectory}. The origin is towards R.A. 
$49.4 \pm 4.1^{\circ}$ and declination $11.2 \pm 1.8^{\circ}$. The heliocentric incoming velocity at infinity of the meteor in right-handed Galactic coordinates is $v_{\infty}\mathrm{(U, V, W) = (32.7 \pm 5.8, -4.5 \pm 1.5, 26.1 \pm 2.0)\;}$ $\mathrm{km\;s^{-1}}$, which is $58\pm6\; \mathrm{km\;s^{-1}}$ away from the velocity of the Local Standard of Rest (LSR), $\mathrm{(U, V, W)_{LSR}}$ $= (-11.1, -12.2, -7.3)\;$ $\mathrm{km\;s^{-1}}$ \citep{Schonrich2010}.

\subsection{Size distribution}
Given the impact speed of the meteor, $\sim 44.8 \; \mathrm{km\;s^{-1}}$, and the total impact energy, $4.6 \times 10^{18}$ ergs, the meteor mass was approximately $4.6 \times 10^5 $ g. Assuming bulk density values of 1.7 $\mathrm{g/cm^{3}}$ and 0.9 $\mathrm{g/cm^{3}}$ for Type II and Type IIIa objects respectively, we obtain a radius, $R$, of 0.4m -- 0.5m for a spherical geometry \citep{Ceplecha1988, Palotai2018}.

The CNEOS catalog includes bolide events at a relatively high frequency for the past decade, so we approximate the yearly detection rate of interstellar meteors to be at least $\sim 0.1 \mathrm{\: yr^{-1}}$. We estimate the number density of similarly sized interstellar objects by dividing the yearly detection rate by the product of the impact speed of the meteor and the cross sectional area of the Earth, finding the approximate number density of interstellar objects with a size of order $R\sim0.45$m and a speed $v \sim 60\; \mathrm{km\;s^{-1}} \; \mathrm{km \; s^{-1}}$ relative to the LSR, to be, 

\begin{equation}
n\sim\frac{0.1 \: \mathrm{yr^{-1}}}{(13 \: \mathrm{ AU/yr})(5.7 \times 10^{-9} \: \mathrm{AU^2})} \sim 10^6 \: \mathrm{ AU^{-3}}.
\end{equation}

Given 95\% Poisson uncertainties, the inferred\footnote{Gravitational focusing by the Earth is negligible since the meteor speed exceeds considerably the escape speed from the Earth. The density enhancement due to gravitational focusing by the Sun is well below the uncertainty in the estimated value of $n$, so that our inferred range of local values also corresponds to the density outside of the Solar System.} local number density for interstellar objects of this size is $n = 10^{6{^{+0.75}_{-1.5}}} \; \mathrm{AU^{-3}}$. This figure necessitates $6 \times 10^{22{^{+0.75}_{-1.5}}}$ similarly size objects, or 0.2 -- 20 Earth masses of material, to be ejected per local star. This is at tension with the fact that a minimum-mass solar nebula is expected to have about an Earth mass of total planetesimal material interior to the radius where the orbital speed is $\sim 60 \; \mathrm{km\;s^{-1}}$ \citep{Desch2007}, with similar values for other planetary systems \citep{Kuchner2004}. Our inferred abundance for interstellar meteors should be viewed as a lower limit since the CNEOS data might have a bias against detection of faster meteors \citep{Brown2016}.

\begin{figure}
  \centering
  \includegraphics[width=\linewidth]{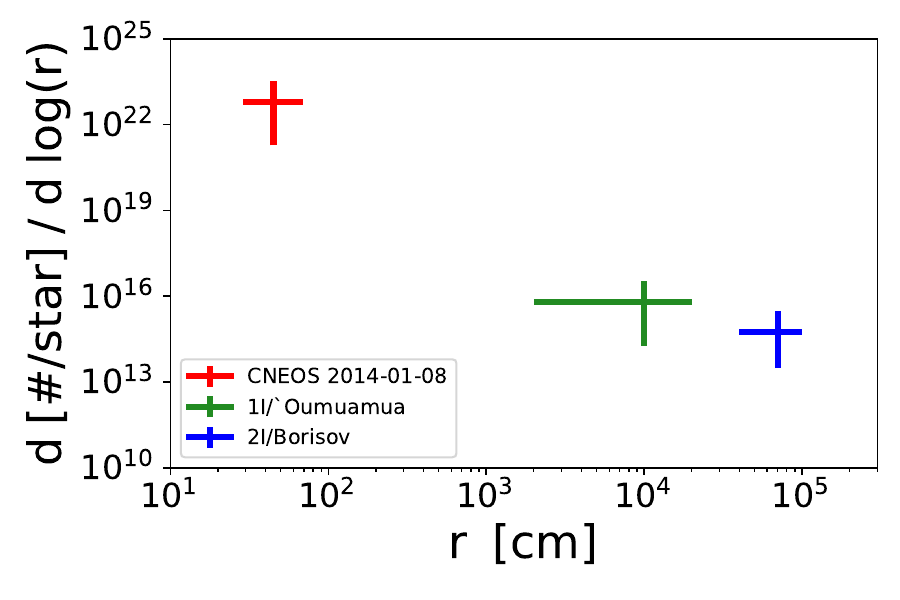}
    \caption{CNEOS 2014-01-08-like objects, `Oumuamua-like objects, and Borisov-like objects in size-abundance parameter space, expressed as number per star per differential unit of log size. The vertical error bars correspond to the 95\% Poisson uncertainties, while the horizontal error bars for `Oumuamua and Borisov correspond to the discrete ranges discussed in the text.}
    \label{fig:sizedist}
\end{figure}

CNEOS 2014-01-08, `Oumuamua, and Borisov together serve as important calibration points for the size distribution of interstellar objects. Figure \ref{fig:sizedist} illustrates the three classes of interstellar objects considered here in size-abundance parameter space. We use a Monte Carlo simulation to characterize the slope of this size distribution. For each run of the simulation, we select a size and a population abundance for the three interstellar objects. The sizes for the CNEOS 2014-01-08, `Oumuamua, and Borisov are drawn from the triangular distributions spanning  $0.3 - 0.7 \mathrm{\; m}$,\footnote{Encapsulating a factor of 3 uncertainty in density in both directions, as well as $\pm20\%$ uncertainty in the velocity, to account for measurement error.} $20 - 200 \mathrm{\; m}$ \citep{Trilling2018}, and $0.4 - 1 \mathrm{\; km}$ \citep{2020ApJ...888L..23J, 2021MNRAS.507L..16S}, respectively, with peaks at $0.45 \mathrm{\; m}$, $100 \mathrm{\; m}$, and $0.7 \mathrm{\; km}$. The abundances are drawn from the Poisson distributions for sample size of one with central values of $10^{6} \mathrm{\; AU}$, $10^{-1} \mathrm{\; AU}$, and $9 \times 10^{-3} \mathrm{\; AU}$, respectively. Given the three size-abundance tuples, each run of the simulation fits a linear least square to the log-log transformed values, resulting in a power-law size distribution fit with normalization $k$ and slope $q$, corresponding to a size distribution with the form $N(>R) \sim k R^{1-q}$. We repeat this routine $10^5$ times, thereby constructing a distribution of $q$ corresponding to the uncertainties in CNEOS 2014-01-08-like, `Oumuamua-like, and Borisov-like populations. Our model applies if `Oumuamua was not primarily composed of hydrogen or helium.

We find that, over the size range from CNEOS 2014-01-08 (m-scale) to Borisov (km-scale), the best fit slope is $q \approx 3.6 \pm 0.5$, where the quoted uncertainty corresponds to $2 \sigma$. The result is in good agreement with $q = 3.5$ the analytical collisional model by \cite{Dohnanyi1969},  but is not inconsistent with the $q = 4$ scale-free power law model, which contains equal mass per logarithmic bin. The power-law extrapolation may not hold for all bolide radii down to dust particles.


\section{Detection Strategy}
The cores of meteoroids with radii larger than $\sim 5$cm can reach the ground in the form of meteorites \citep{Kruger2014}. Additionally, meteoroids on smaller size scales could be accelerated from the Poynting-Robertson effect and could have potential origins in the interstellar medium. Hence, interstellar meteors above this size are optimal for a systematic study of physical extrasolar material (in addition to the spectroscopy of the hot gases as the meteor burns up). Since we expect interstellar meteors of this size to strike the Earth at least a few times per year, a network of all-sky camera systems monitoring the sky above all land on Earth could detect an interstellar meteor of this size every few years. Such detections can be made with science-grade video cameras, such as those used in AMOS, CAMO, and CAMS\footnote{http://cams.seti.org/} \citep{Toth2015, Weryk2013}.

A conservative estimate for the total area of $\sim70$km altitude atmosphere visible from a system of two all-sky camera systems separated by 100km is $5 \times 10^5 \; \mathrm{km^2}$, so to cover all land on Earth would require $\sim 300$ systems, or $\sim 600$ total all-sky camera systems, similar to CAMO but with an all-sky field of view, like AMOS and CAMS \citep{Weryk2013, Toth2015}.

We therefore advocate for a network of all-sky camera systems to conduct real-time remote spectroscopy of the hot gases as $r \geq 5 \times 10^{-2}\; \mathrm{m}$ interstellar meteors burn up, and to precisely determine their trajectories for the immediate retrieval of interstellar meteorite samples.

\section{Discussion}
\label{sec:discussion}
We presented and analyzed impact data from the meteor detected at 2014-01-08 17:05:34 UTC, showing that it originated from an unbound hyperbolic orbit with 99.999\% confidence. The U.S. Department of Defense has since verified in a letter referencing this work that ``the velocity estimate reported to NASA is sufficiently accurate to indicate an interstellar trajectory'' \citep{Shaw2022}. The meteor had an asymptotic speed of $v_{\infty} \sim 42.1 \pm 5.5\; \mathrm{km \; s^{-1}}$ outside of the Solar System. Its size, trajectory, and excess speed exclude the possibility that it was gravitationally scattered within the Solar System prior to impact \citep{Wiegert2014}. Its $\sim 58\pm6\; \mathrm{km \; s^{-1}}$ deviation from the LSR suggests that it perhaps originated in the thick disk, which has velocity dispersion components of $\mathrm{(\sigma_U, \sigma_V, \sigma_W) = (50, 50, 50)\;} \mathrm{km \; s^{-1}}$  relative to the LSR \citep{Bland-Hawthorn2016}. However, the ratio of local thick disk stars to thin disk stars is 0.04, making this a minority population. Alternatively, for a parent planetary system with a more typical velocity relative to the LSR, the object could have originated in the deep interior, where the orbital speeds of objects are of the necessary magnitude. Either way, the meteor had an unusual origin.

The discovery of additional interstellar meteors will serve as an important calibration for population-wide parameters of interstellar objects, including their abundance and origin.

We estimate the impact rate of similarly sized objects with the Earth, given 95\% Poisson distribution confidence intervals, to be at least $0.1\substack{+0.457 \\ -0.097}$ per year. Future meteor surveys could flag incoming objects with excess heliocentric velocities for follow-up pre-impact observations. Spectroscopy of gaseous debris from these objects as they burn up in the Earth's atmosphere would reveal their composition. Given that some isotope ratios are expected to be markedly different for objects of interstellar origin compared to the Solar System, could validate an interstellar origin \citep{Lingam2018, Forbes2019}. Precision tracking with the upcoming Large Synoptic Survey Telescope (LSST\footnote{https://www.lsst.org/}) could determine the trajectory of meteors of interstellar origin to their parent systems in the Gaia catalog.\footnote{https://gea.esac.esa.int/archive/} Our discovery also implies that at least $4.5 \times 10^{8{^{+0.75}_{-1.5}}}$ similarly sized interstellar bolide events have occurred over Earth's lifetime. Potentially, interstellar meteors could deliver life from another planetary system and mediate panspermia \citep{Ginsburg2018}. Interestingly, the high speed for the meteor discussed here implies a likely origin in the habitable zone of the abundant population of dwarf stars, indicating that similar objects could carry life from their parent planetary systems.

We also analyzed the updated size distribution of interstellar meteors, deriving a range of possible slopes of $q \sim 3.6 \pm 0.5$, consistent with the analytical model by \cite{Dohnanyi1969}. Finally, we presented a strategy for studying interstellar meteors: using a network of $\sim 600$ all-sky camera systems to determine the orbits and chemical compositions of $r \geq 5$cm meteors. This method also allows for the possibility of retrieving interstellar meteorite samples. By extrapolating the trajectory of each meteor backward in time and analyzing the relative abundances of each meteor's chemical isotopes, one can match meteors to their parent stars and reveal insights into planetary system formation. R-processed elements, such as Eu, can be detected in the atmospheres of stars \citep{Frebel2016}, so their abundances in meteor spectra can serve as important links to parent stars. This new field of astronomical research is significant as it would save the trip and allow us to study samples of materials from other planetary systems, be it natural or artificial in origin.
\section*{Acknowledgements}
\vspace{0.1in} 

We thank Matt Daniels for his exceptional help in procuring official confirmation from the U.S. government that the measurement uncertainties of the 2014-01-08 meteor were small enough to confirm its interstellar nature. We thank Pete Worden for introducing the authors to Matt Daniels. We thank Lt. Gen. John Shaw, Joel Mozer, and an anonymous analyst, for their detailed examination of the relevant Department of Defense data that led to the confirmation of the meteor's interstellar origin. We thank Alan Hurd and Matthew Heavner for initiating the search for clarification regarding the quality of the data for the 2014-01-08, and for retrieving upper bounds on the velocity measurement uncertainties. Without the help of all aforementioned individuals, the interstellar origin of this meteor would have been enshrouded in uncertainty. We also thank Nathan Golovich, Jeff Kuhn, Manasvi Lingam, Matthew Payne, Steinn Sigurdsson, and Peter Veres for helpful comments on the manuscript. This work was supported in part by a grant from the Breakthrough Prize Foundation. 


\begin{thebibliography}{}
\makeatletter
\relax
\def\mn@urlcharsother{\let\do\@makeother \do\$\do\&\do\#\do\^\do\_\do\%\do\~}
\def\mn@doi{\begingroup\mn@urlcharsother \@ifnextchar [ {\mn@doi@}
  {\mn@doi@[]}}
\def\mn@doi@[#1]#2{\def\@tempa{#1}\ifx\@tempa\@empty \href
  {http://dx.doi.org/#2} {doi:#2}\else \href {http://dx.doi.org/#2} {#1}\fi
  \endgroup}
\def\mn@eprint#1#2{\mn@eprint@#1:#2::\@nil}
\def\mn@eprint@arXiv#1{\href {http://arxiv.org/abs/#1} {{\tt arXiv:#1}}}
\def\mn@eprint@dblp#1{\href {http://dblp.uni-trier.de/rec/bibtex/#1.xml}
  {dblp:#1}}
\def\mn@eprint@#1:#2:#3:#4\@nil{\def\@tempa {#1}\def\@tempb {#2}\def\@tempc
  {#3}\ifx \@tempc \@empty \let \@tempc \@tempb \let \@tempb \@tempa \fi \ifx
  \@tempb \@empty \def\@tempb {arXiv}\fi \@ifundefined
  {mn@eprint@\@tempb}{\@tempb:\@tempc}{\expandafter \expandafter \csname
  mn@eprint@\@tempb\endcsname \expandafter{\@tempc}}}

\bibitem[\protect\citeauthoryear{Afanasiev, Kalenichenko \& Karachentsev}{Afanasiev et~al.}{2007}]{Afanasiev2007}
Afanasiev V. L., Kalenichenko V. V., Karachentsev I. D., 2007, \mn@doi [Astrophysical Bulletin] {10.1134/S1990341307040013}, 62(4), 319


\bibitem[\protect\citeauthoryear{Baggaley et~al.}{Baggaley et~al.}{1993}]{Baggaley1993}
Baggaley W. J., Taylor, D.A. \& Steel, I.D. 1993,
Meteoroids and their Parent Bodies, Proc. Int. Astron. Symp., 53

\bibitem[\protect\citeauthoryear{Baggaley}{Baggaley}{2000}]{Baggaley2000}
Baggaley W. J., 2000, \mn@doi [Journal of Geophysical Research] {10.1029/1999JA900383}, 105(A5), 10353

\bibitem[\protect\citeauthoryear{Bannister et~al.,}{Bannister et~al.}{2017}]{Bannister2017}
Bannister M.~T.,  et~al., 2017, \mn@doi [The Astrophysical Journal]
  {10.3847/2041-8213/aaa07c}, 851, L38
  
\bibitem[\protect\citeauthoryear{Bialy \& Loeb}{Bialy \& Loeb}{2018}]{Bialy2018}
Bialy S.,  Loeb A.,  2018, \mn@doi [The Astrophysical Journal]
  {10.3847/2041-8213/aaeda8}, 868, L1
 
\bibitem[\protect\citeauthoryear{Bland-Hawthorn \& Gerhard}{Bland-Hawthorn \&
  Gerhard}{2016}]{Bland-Hawthorn2016}
Bland-Hawthorn J.,  Gerhard O.,  2016, \mn@doi [Annual Review of Astronomy and
  Astrophysics] {10.1146/annurev-astro-081915-023441}, 54, 529
  
\bibitem[\protect\citeauthoryear{Bolin et~al.,}{Bolin et~al.}{2017}]{Bolin2017}
Bolin B.~T.,  et~al., 2017, \mn@doi [The Astrophysical Journal Letters, Volume
  852, Issue 1, article id. L2, 10 pp. (2018).] {10.3847/2041-8213/aaa0c9}, 852


\bibitem[\protect\citeauthoryear{Brown et~al.,}{Brown et~al.}{2016}]{Brown2016}
Brown P., et~al., 2016, \mn@doi [Icarus]
  {10.1016/j.icarus.2015.11.022}, 266, 96
  
\bibitem[\protect\citeauthoryear{Ceplecha}{Ceplecha}{1988}]{Ceplecha1988}
Ceplecha Z.,  1988, Bulletin of the Astronomical Institutes of Czechoslovakia, 39, 221

\bibitem[\protect\citeauthoryear{Collins, Melosh \& Marcus}{Collins et~al.}{2005}]{Collins2005}
Collins G. S., Melosh J. H., Marcus, R. A., 2005, \mn@doi [Meteoritics \& Planetary Science] {10.1111/j.1945-5100.2005.tb00157.x}, 40, 6, 817

\bibitem[\protect\citeauthoryear{Desch}{Desch}{2007}]{Desch2007}
Desch S.J., 2007, \mn@doi [The Astrophysical Journal]
  {10.1086/522825}, 671(1), 878
  
\bibitem[\protect\citeauthoryear{Dohnanyi}{Dohnanyi}{1969}]{Dohnanyi1969}
Dohnanyi J. S., 1969, \mn@doi [Journal of Geophysical Research]
  {10.1029/JB074i010p02531}, 74, 2531
  
\bibitem[\protect\citeauthoryear{Devillepoix et~al.,}{Devillepoix et~al.}{2019}]{Devillepoix2019}
Devillepoix H.A.R., et~al., 2019, \mn@doi [Monthly Notices of the Royal Astronomical Society]
  {10.1093/mnras/sty3442}, 483(4), 5166
  
\bibitem[\protect\citeauthoryear{Engelhardt et~al.,}{Engelhardt et~al.}{2017}]{Engelhardt2017}
Engelhardt T., et~al., 2017, \mn@doi [The Astronomical Journal] {10.3847/1538-3881/aa5c8a}, 153, 133
  
\bibitem[\protect\citeauthoryear{Fitzsimmons et~al.,}{Fitzsimmons
  et~al.}{2018}]{Fitzsimmons2018}
Fitzsimmons A.,  et~al., 2018, \mn@doi [Nature Astronomy]
  {10.1038/S41550-017-0361-4}, 2, 133

\bibitem[\protect\citeauthoryear{Forbes \& Loeb}{Forbes \& Loeb}{2019}]{Forbes2019}
Forbes J.,  Loeb A.,  2018, {Turning up the heat on `Oumuamua}.
 (\mn@eprint {arXiv} {1901.00508})
 
\bibitem[\protect\citeauthoryear{Frebel, Yu \& Jacobson}{Frebel et~al.}{2016}]{Frebel2016}
Frebel A., Yu Q., Jacobson H. R., 2016, \mn@doi [J. Phys.: Conf. Ser.] {10.1088/1742-6596/665/1/012051}, 665 012051

\bibitem[\protect\citeauthoryear{Froncisz, Brown, \& Weryk}{2020}]{2020P&SS..19004980F} Froncisz M., Brown P., Weryk R.~J., 2020, \mn@doi [P\&SS] {10.1016/j.pss.2020.104980}, 190, 104980

\bibitem[\protect\citeauthoryear{Gaidos, Williams  \& Kraus}{Gaidos
  et~al.}{2017}]{Gaidos2017}
Gaidos E.,  Williams J.,   Kraus A.,  2017, \mn@doi [Research Notes of the AAS]
  {10.3847/2515-5172/aa9851}, 1, 13

\bibitem[\protect\citeauthoryear{Ginsburg, Lingam \& Loeb}{Ginsburg et~al.}{2018}]{Ginsburg2018}
Ginsburg I., Lingam M., Loeb A., 2018, \mn@doi [The Astrophysical Journal] {10.3847/2041-8213/aaef2d}, 868(1), L12

\bibitem[\protect\citeauthoryear{Granvik \& Brown}{Granvik \& Brown}{2018}]{Granvik2018}
Granvik M., Brown P., 2018, \mn@doi [Icarus] {10.1016/j.icarus.2018.04.012}, 311(1), 271

\bibitem[\protect\citeauthoryear{Grun et~al.}{Grun et~al.}{2000}]{Grun2000}
Grun E., et~al., 2000, J. Geophys. Res., 103

\bibitem[\protect\citeauthoryear{Guzik et al.}{2020}]{2020NatAs...4...53G} Guzik P., Drahus M., Rusek K., Waniak W., Cannizzaro G., Pastor-Marazuela I., 2020, NatAs, 4, 53

\bibitem[\protect\citeauthoryear{Hajdukova}{Hajdukova}{1994}]{Hajdukova1994}
Hajdukova M., Jr., 1994, \mn@doi [Astronomy and
  Astrophysics], 288(1), 330
  
\bibitem[\protect\citeauthoryear{Hajdukova \& Paulech}{Hajdukova \& Paulech}{2002}]{Hajdukova2002}
Hajdukova M., Jr., Paulech, T., 2002, Asteroids, Comets, and Meteors: ACM 2002, ed. B. Warmbein (Noordwijk: ESA), 173, ESA SP-500
  
\bibitem[\protect\citeauthoryear{Hajdukova, Sterken \& Wiegert}{Hajdukova et~al.}{2018}]{Hajdukova2018}
Hajdukova M., Sterken, V., Wiegert, P., 2018, \mn@doi [European Planetary Science Congress], 12

\bibitem[\protect\citeauthoryear{Hawkes \& Woodworth}{Hawkes \& Woodworth}{1999}]{Hawkes1999}
Hawkes, R. L., Woodworth, S. C., 1997, J. R. Astron. Soc. Canada, 91, 218

\bibitem[\protect\citeauthoryear{Hoang, Loeb, Lazarian \& Cho}{Hoang et~al.}{2018}]{Hoang2018}
Hoang T.,  Loeb A.,  Lazarian A.,  Cho J., 2018, \mn@doi [The Astrophysical Journal]
  {10.3847/1538-4357/aac3db}, 860(1), 42

\bibitem[\protect\citeauthoryear{Jenniskens}{Jenniskens}{2015}]{Jenniskens2015}
Jenniskens P.,  2015, Encyclopedia of Atmospheric Sciences (Second Edition)

\bibitem[\protect\citeauthoryear{Jewitt, Luu, Rajagopal, Kotulla, Ridgway, Liu
  \& Augusteijn}{Jewitt et~al.}{2017}]{Jewitt2017}
Jewitt D.,  Luu J.,  Rajagopal J.,  Kotulla R.,  Ridgway S.,  Liu W.,
  Augusteijn T.,  2017, \mn@doi [The Astrophysical Journal]
  {10.3847/2041-8213/aa9b2f}, 850, L36

\bibitem[\protect\citeauthoryear{Jewitt et al.}{2020}]{2020ApJ...888L..23J} Jewitt D., Hui M.-T., Kim Y., Mutchler M., Weaver H., Agarwal J., 2020, ApJL, 888, L23

\bibitem[\protect\citeauthoryear{Kuchner}{Kuchner}{2004}]{Kuchner2004}
Kuchner M.J., 2004, \mn@doi [The Astrophysical Journal] {10.1086/422577}, 612, 1147

\bibitem[\protect\citeauthoryear{Kruger \& Grun}{Kruger \& Grun}{2014}]{Kruger2014}
Kruger H.,  Grun E.,  2014, Encyclopedia of the Solar System (Third Edition)

\bibitem[\protect\citeauthoryear{Landgraf \& Grun}{Landgraf \& Grun}{1998}]{Landgraf1998}
Landgraf, M., Grun, E., 1998, The Local Bubble and Beyond (Lecture Notes in Physics, IAU Colloq. 166), Vol. 506, ed. D. Breitschwerdt, M. J. Freyberg, \& J. Truempe (Berlin: Springer), 381
  
\bibitem[\protect\citeauthoryear{Landgraf et~al.}{Landgraf et~al.}{2000}]{Landgraf2000}
Landgraf, M., Baggaley, W. J., Grun, E., Kruger, H., Linkert, G., 2000, J. Geophys. Res., 105, 10343

\bibitem[\protect\citeauthoryear{Lingam \& Loeb}{Lingam \& Loeb}{2018a}]{Lingam2018}
Lingam M.,  Loeb A.,  2018a, \mn@doi [The Astronomical Journal]
  {10.3847/1538-3881/aae09a}, 156

\bibitem[\protect\citeauthoryear{Mamajek}{Mamajek}{2017}]{Mamajek2017}
Mamajek E.,  2017, \mn@doi [Research Notes of the AAS]
  {10.3847/2515-5172/aa9bdc}, 1, 21

\bibitem[\protect\citeauthoryear{Mathews et~al.}{Mathews et~al.}{1998}]{Mathews1998}
Mathews, D. J., Meisel, D. D., Janches, D., Getman, S. V., Zhou, Q.-H.,  1998, Meteoroids 1998 (Proc. Int. Conf.), ed. W. J. Baggaley \& V. Porubcan (Bratislava: Astronomical Institute of the Slovak Academy of Sciences), 79

\bibitem[\protect\citeauthoryear{Mathis, Rumpl \& Nordsieck}{Mathis et~al.}{1977}]{Mathis1977}
Mathis, J. S., Rumpl, W., \& Nordsieck, K. H., 1977, The Astrophysical Journal, 217, 425

\bibitem[\protect\citeauthoryear{Meech et~al.,}{Meech et~al.}{2017}]{Meech2017}
Meech K.~J.,  et~al., 2017, \mn@doi [Nature] {10.1038/nature25020}, 552, 378

\bibitem[\protect\citeauthoryear{Meisel, Janches  \& Matthews}{Meisel et~al.}{2002a}]{Meisel2002a}
Meisel D. D.,  Janches D.,   Mathews J. D.,  2002a, \mn@doi [The Astrophysical Journal]
  {10.1086/322317}, 567, 323

\bibitem[\protect\citeauthoryear{Meisel, Janches  \& Matthews}{Meisel et~al.}{2002b}]{Meisel2002b}
Meisel D. D.,  Janches D.,   Mathews J. D.,  2002b, \mn@doi [The Astrophysical Journal]
  {10.1086/342919}, 567, 323
  
\bibitem[\protect\citeauthoryear{Micheli et~al.,}{Micheli
  et~al.}{2018}]{Micheli2018}
Micheli M.,  et~al., 2018, \mn@doi [Nature] {10.1038/s41586-018-0254-4}, 559, 223

\bibitem[\protect\citeauthoryear{Musci et~al.,}{Musci et~al.}{2012}]{Musci2012}
Musci R.,  et~al., 2012, \mn@doi [The Astrophysical Journal] {10.1088/0004-637X/745/2/161}, 745, 161

\bibitem[\protect\citeauthoryear{Opik}{Opik}{1950}]{Opik1950}
Opik, E. J., 1950, \mn@doi [Irish Astronomical Journal], 1(3), 80

\bibitem[\protect\citeauthoryear{Palotai et al.}{Palotai et al.}{2018}]{Palotai2018}
Palotai, C., et al.,  2018, {Analysis of June 2, 2016 bolide event}.
 (\mn@eprint {arXiv} {1801.05072})
 
\bibitem[\protect\citeauthoryear{Rein \& Liu}{Rein \& Liu}{2012}]{Rein2012}
Rein H.,  Liu S.-F.,  2012, \mn@doi [Astronomy {\&} Astrophysics]
  {10.1051/0004-6361/201118085}, 537, A128

\bibitem[\protect\citeauthoryear{Rein \& Spiegel}{Rein \&
  Spiegel}{2014}]{Rein2014}
Rein H.,  Spiegel D.S.,  2014, \mn@doi [Monthly Notices of the Royal Astronomical
  Society] {10.1093/mnras/stu2164}, 446(2), 1424

\bibitem[\protect\citeauthoryear{Schonrich, Binney, \& Dehnen}{Schonrich et al.}{2010}]{Schonrich2010}
Schonrich, R., Binney, J., \& Dehnen, W.,  2010, \mn@doi [Monthly Notices of the Royal Astronomical
  Society] {10.1111/j.1365-2966.2010.16253.x}, 403(4), 1829

\bibitem[\protect\citeauthoryear{Seligman, Luaghlin
  \& Batygin}{Seligman et~al.}{2019}]{Seligman2019}
Seligman D.,  Laughlin G.,  Batygin K., 2019, \mn@doi [The Astrophysical Journal Letters]
  {arXiv:1903.04723}, 

\bibitem[\protect\citeauthoryear{Shaw}{Shaw}{2022}]{Shaw2022}
Shaw, J. E., 2022, U.S. Department of Defense, Confirmation of Interstellar Object, \url{https://twitter.com/US_SpaceCom/status/1511856370756177921?s=20&t=vzmmhRf6WthBaCt7BGzReA}

\bibitem[\protect\citeauthoryear{Siraj \&
  Loeb}{Siraj \&
  Loeb}{2019a}]{Siraj2019a}
Siraj A. \& Loeb A., 2019, \mn@doi [The Astrophysical Journal]{10.3847/2041-8213/ab042a}, 872(1), L10

\bibitem[\protect\citeauthoryear{Siraj \&
  Loeb}{Siraj \&
  Loeb}{2019b}]{Siraj2019b}
Siraj A. \& Loeb A., 2019, \mn@doi [Research Notes of the American Astronomical Society]{10.3847/2515-5172/aafe7c}, 3(1), 15

\bibitem[\protect\citeauthoryear{Siraj \& Loeb}{2021}]{2021MNRAS.507L..16S} Siraj A., Loeb A., 2021, MNRAS, 507, L16

\bibitem[\protect\citeauthoryear{Taylor, Baggaley \& Steel}{Taylor et~al.}{1996}]{Taylor1996}
Taylor A. D., Baggaley W. J., Steel D. I., 2018, \mn@doi [Nature] {10.1038/380323a0}, 380, 323

\bibitem[\protect\citeauthoryear{Toth}{Toth et~al.}{2015}]{Toth2015}
Toth J., et~al., 2015, \mn@doi [Planetary \& Space Science] {10.1016/j.pss.2015.07.007}, 118, 102

\bibitem[\protect\citeauthoryear{Trilling et al.}{Trilling et al.}{2018}]{Trilling2018}
Trilling, D., et al.,  2018, \mn@doi [The Astronomical Journal] {	10.3847/1538-3881/aae88f},
  156, 261.
  
\bibitem[\protect\citeauthoryear{Weryk \& Brown}{Weryk \& Brown}{2004}]{Weryk2004}
Weryk R. J., Brown P., 2004, \mn@doi [Icarus] {10.1007/s11038-005-9034-x},
  95, 221.
  
\bibitem[\protect\citeauthoryear{Weryk}{Weryk et~al.}{2013}]{Weryk2013}
Weryk R. J., et al., 2013, \mn@doi [Icarus] {10.1016/j.icarus.2013.04.025},
  225, 614.

\bibitem[\protect\citeauthoryear{Wiegert}{Wiegert}{2014}]{Wiegert2014}
Wiegert P.,  2014, \mn@doi [Icarus] {10.1016/j.icarus.2014.06.031},
  242, 112.
  
\bibitem[\protect\citeauthoryear{Ye, Zhang, Kelley  \& Brown}{Ye
  et~al.}{2017}]{Ye2017}
Ye Q.-Z.,  Zhang Q.,  Kelley M. S.~P.,   Brown P.~G.,  2017, \mn@doi [The
  Astrophysical Journal] {10.3847/2041-8213/aa9a34}, 851, L5

\bibitem[\protect\citeauthoryear{Zuluaga}{Zuluaga}{2019}]{Zuluaga2019}
Zuluaga J. I,  2019, \mn@doi [Research Notes of the AAS]
  {10.3847/2515-5172/ab1de3}, 3, 68
  
\makeatother
\end{thebibliography}
\end{document}